# Theoretical description of anomalous properties of novel room temperature multiferroics $Pb(Fe_{1/2}Ta_{1/2})_x(Zr_{0.53}Ti_{0.47})_{1-x}O_3$ and $Pb(Fe_{1/2}Nb_{1/2})_x(Zr_{0.53}Ti_{0.47})_{1-x}O_3$

Maya D. Glinchuk[1], Eugene A. Eliseev[1], and Anna N. Morozovska[2*]

[1] Institute for Problems of Materials Science, National Academy of Sciences of Ukraine,
Krjijanovskogo 3, 03142 Kyiv, Ukraine

[2] Institute of Physics, National Academy of Sciences of Ukraine,
46, pr. Nauky, 03028 Kyiv, Ukraine

**Abstract**

The theoretical description of the anomalous ferroelectric, ferromagnetic and magnetoelectric properties of $Pb(Fe_{1/2}Ta_{1/2})_x(Zr_{0.53}Ti_{0.47})_{1-x}O_3$ (PFTx-PZT(1-x)) and $Pb(Fe_{1/2}Nb_{1/2})_x(Zr_{0.53}Ti_{0.47})_{1-x}O_3$ (PFNx-PZT(1-x)) micro-ceramics is given for the first time. We performed calculations of temperature, composition and external field dependence of ferroelectric, ferromagnetic and antiferromagnetic phases transition temperatures, remanent polarization, magnetization, hysteresis loops, coercive fields, dielectric permittivity and magnetoelectric coupling. Special attention was paid to comparison of the developed theory with experiments. It appeared possible to describe adequately the main experimental results including a reasonable agreement between the shape of calculated hysteresis loops and remnant polarization value with measured loops and polarization. Information about linear and nonlinear magnetoelectric coupling coefficients was extracted from the experimental data. From the fitting of experimental data with theoretical formula it appeared possible to obtain composition dependence of Curie-Weiss constant that is known to be inversely proportional to harmonic (linear) dielectric stiffness, as well as the strong nonlinear dependence of anharmonic parameters of free energy. Keeping in mind the essential influence of these parameters on the multiferroic properties the obtained results open the way to govern practically all the material properties with the help of suitable choice of composition. The forecast of the strong enough influence of antiferrodistortive order parameter on the transition temperatures and so on the phase diagrams and properties of multiferroics is made on the basis of the developed theory.

---

[*] Corresponding author: anna.n.morozovska@gmail.com

# I. Introduction

Recently a new hot scientific topic appeared due to the inventing new room-temperature magnetoelectric materials consisting of single-phase multiferroics containing two or more ferroic order parameters, e.g. ferroelectric and ferromagnetic ones. A strong magnetoelectric (ME) coupling existing at room temperature is especially vital for novel functional devices fabrication [1, 2, 3, 4, 5].

Nowadays the attention of researchers and engineers are paid mainly to the solid solutions of ferroelectric antiferromagnets $Pb(Fe_{1/2}Ta_{1/2})O_3$ (PFT) and $Pb(Fe_{1/2}Nb_{1/2})O_3$ (PFN) with $Pb(Zr_{0.53}Ti_{0.47})O_3$ (PZT) [6, 7, 8, 9, 10, 11], further denoted as $Pb(Fe_{1/2}Ta_{1/2})_x(Zr_{0.53}Ti_{0.47})_{1-x}O_3$ (**PFTx-PZT(1-x)**) and $Pb(Fe_{1/2}Nb_{1/2})_x(Zr_{0.53}Ti_{0.47})_{1-x}O_3$ (**PFNx-PZT(1-x)**)

For PFN the antiferromagnetic Neel transition temperature $T_N$ and ferroelectric Curie temperature $T_C$ are $T_N$ = 143–170 K [12, 13, 14] and $T_C$ = 379–393 K [13, 14, 15, 16, 17], while $T_N$ = 133–180 K [13, 18, 19] and $T_C \cong 250$ K [20] for PFT. The second component of the solid solutions is well known ferroelectric with $T_C$ varied in the range 666–690 K and high piezoelectric effect [21].

Sanchez et al successfully prepared single-phase PFTx-PZT(1-x) (x = 0.3 – 0.4) ceramics by conventional solid-state route and performed temperature-dependent XRD studies [6, 9] and resonant ultrasonic spectroscopy [11], which revealed structural phases with orthorhombic, tetragonal and cubic symmetries under heating. In particular their samples show several sequential structural phase transitions from cubic to tetragonal phase at about 1123 K, from the tetragonal to orthorhombic at 520 K for x = 0.3 and at 475 K for x = 0.4; and then from the orthorhombic to rhombohedral at 230 K for x = 0.3 and at 270 K for x = 0.4. The sequence is similar to that of ferroelectric phases in barium titanate. PFTx-PZT(1-x) showed both ferroelectric and ferromagnetic ordering at room an higher temperatures with perfect square hysteresis loop at 550 K. The electric leakage is very low entire the temperature range. Ferromagnetic and magnetoelectric properties of PFTx-PZT(1-x) and PFNx-PZT(1-x) were studied at compositions x=0.1 – 0.4 [6, 8, 9]. At x=0.1 the ferromagnetism is faint, while at x=0.2 – 0.4 PFTx-PZT(1-x) exhibits saturated square-like magnetic hysteresis loops with magnetization 0.1 emu/g at 295 K and pronounced saturated ferroelectric hysteresis with saturation polarization 25 μC/cm$^2$, which actually increases to 40 μC/cm$^2$ in the high temperature tetragonal phase, representing an exciting new room temperature oxide multiferroic [8, 9]. Giant effective ME coefficient of PFTx-PZT(1-x) was reported as 1.3×10$^{-7}$ s/m for x=0.4 [8], however it appeared to be a *nonlinear* effect. Note that theoretical consideration [22] had shown that giant linear ME coefficient is also possible due to the size effect in nanostructure PFT-PZT lamellas.

PFNx-PZT(1-x) demonstrates magnetization loop vs. applied magnetic field at room temperature for the composition range x between 0.1 and 0.4; an improvement in ferromagnetic properties was observed for x=0.2 and x=0.3, while a notable deterioration of the these properties was observed for x=0.1 and 0.4. [6, 9]. Saturated and low loss ferroelectric hysteresis curves with a remanent polarization Ф about 20-30 μC/cm$^2$ was observed in Refs.[6, 9, 23].

Previously we proposed a theoretical description of the nanostructured PFNx-PZT(1-x) and PFTx-PZT(1-x) intriguing phase diagrams and ferromagnetic properties [22]. In particular it was shown that their nanostructure plays a decisive role in the strong ME coupling and the solid solution second component PZT induces ferromagnetic phase by transforming the negative value of magnetic Curie temperature (that defines the magnetic susceptibility behaviour at $T>T_N$) into a positive one due to the ME coupling. However the theoretical description of the room temperature ferromagnetic, anomalous ferroelectric and magnetoelectric properties of PFTx-PZT(1-x) and PFNx-PZT(1-x) microceramics was absent to date.

Therefore the aim of this study is to propose the theoretical description of the aforementioned properties of microceramics, including room temperature ferromagnetic hysteresis, anomalous ferroelectric and magnetoelectric properties dependence on the composition x. The comparison of the developed theory with available experiments is performed. The obtained results poured light on the physical mechanisms of the properties anomalies and so open the way of creating new room temperature multiferroics for modern electronic technique.

## II. Theoretical formalism

The homogeneous bulk density of Landau-Ginzburg potential is the sum of polarization ($g_P$), antimagnetization ($g_L$), magnetization ($g_M$), elastic ($g_{el}$), structural antiferrodistortive ($g_{AFD}$) and magnetoelectric ($g_{ME}$) parts [22, 24, 25]:

$$G_{PM} = g_P + g_L + g_M + g_{el} + g_{AFD} + g_{ME} \tag{1a}$$

The densities

$$g_P = \frac{\alpha_P}{2} P_i^2 + q_{ijkl}^{(e)} u_{ij} P_k P_l + \frac{\beta_{Pij}}{4} P_i^2 P_j^2 + \frac{\gamma_{Pijkl}}{6} P_i^2 P_j^2 P_k^2 P_l^2 + ..., \tag{1a}$$

$$g_L = \frac{\alpha_L}{2} L_i^2 + q_{ijkl}^{(l)} u_{ij} L_k L_l + \frac{\beta_{Lij}}{4} L_i^2 L_j^2 + \frac{\gamma_{Lijkl}}{6} L_i^2 L_j^2 L_k^2 L_l^2 + ..., \tag{1b}$$

$$g_M = \frac{\alpha_M}{2} M_i^2 + q_{ijkl}^{(m)} u_{ij} M_k M + \frac{\beta_{Mij}}{4} M_i^2 M_j^2 + \frac{\gamma_{Lijkl}}{6} M_i^2 M_j^2 M_k^2 M_l^2 + ..., \tag{1c}$$

$$g_{el} = \frac{c_{ijkl}}{2} u_{ij} u_{kl} + \frac{A_{ijklmn}}{2} u_{ij} u_{kl} P_m P_n + \frac{B_{ijklmn}}{2} u_{ij} u_{kl} L_m L_n + \frac{C_{ijklmn}}{2} u_{ij} u_{kl} M_m M_n, \tag{1d}$$

$$g_{AFD} = \frac{\alpha_\Phi}{2} \Phi_i^2 + q_{ijkl}^{(r)} u_{ij} \Phi_k \Phi_l + \frac{\beta_{\Phi ij}}{4} \Phi_i^2 \Phi_j^2. \tag{1e}$$

Here **P** is the polarization, $L_i = (M_{ai} - M_{bi})/2$ is the components of antimagnetization vector of two equivalent sub-lattices $a$ and $b$, and $M_i = (M_{ai} + M_{bi})/2$ is the magnetization vector components; $\Phi_i$ is the multi-component antiferrodistortive order parameter, $u_{ij}$ is elastic strain tensor; $q_{ijkl}^{(e)}$, $q_{ijkl}^{(l)}$, $q_{ijkl}^{(r)}$ and $q_{ijkl}^{(m)}$ are the bulk electrostriction, antimagnetostriction, rotostriction and magnetostriction coefficients correspondingly, $c_{ijkl}$ are elastic stiffness. Since PZT becomes antiferrodistortive at Zr content more than 50% [21], the AFD contribution has to be taken into account at x<0.5.

The coefficient $\alpha_P$ linearly depends on temperature, i.e. $\alpha_P = \alpha_{TY}(T - T_Y^C)$ [22], where Y="N" or "T"; "N" corresponds to PFN and "T" corresponds to PFT. $T_Y^C$ is the ferroelectric Curie temperature of homogeneous bulk. Similarly, $\alpha_L = \alpha_{LT}(T - T_Y^N)$ and $\alpha_M = \alpha_{MT}(T - \theta_Y^C)$. Antiferromagnetic Neel and magnetic Curie temperatures, $T_Y^N$ and $\theta_Y^C$, correspond to the bulk. Coefficient $\alpha_\Phi = \alpha_{\Phi T}(T - T_Y^\Phi)$, where $T_Y^\Phi$ is the antiferrodistortive transition temperature. Following the logic used for e.g. CaTiO$_3$ [26, 27, 28] any symmetry antiferrodistortive transitions can be also determined from the consideration of $g_{AFD}$ with multi-component order parameter $\Phi_i$. Note, that all the quantities can depend on the composition x of the solid solutions PFNx-PZT(1-x) and PFTx-PZT(1-x).

The linear and quadratic ME energy, that includes rotoelectric, rotomagnetic and rotoantiferromanetic coupling, is:

$$g_{ME} = \begin{pmatrix} \mu_{ij} M_i P_j + \tilde{\mu}_{ij} L_i P_j + \dfrac{\eta_{ijkl}^{FM}}{2} M_i M_j P_k P_l + \dfrac{\eta_{ijkl}^{AFM}}{2} L_i L_j P_k P_l + \\ \dfrac{\eta_{ijkl}^{\Phi P}}{2} \Phi_i \Phi_j P_k P_l + \dfrac{\eta_{ijkl}^{\Phi M}}{2} \Phi_i \Phi_j M_k M_l + \dfrac{\eta_{ijkl}^{\Phi L}}{2} \Phi_i \Phi_j L_k L_l \end{pmatrix}, \quad (2)$$

$\mu_{ij}$ is the bilinear ME coupling term, $\eta_{ijkl}^{FM}$ and $\eta_{ijkl}^{AFM}$ are the components of the biquadratic ME coupling tensor, $\eta_{ijkl}^{\Phi P}$, $\eta_{ijkl}^{\Phi M}$ and $\eta_{ijkl}^{\Phi L}$ are the components of the rotoelectric, rotomagnetic and rotoantiferromanetic coupling tensor. Tensors $\eta_{ijkl}^{FM}$ and $\eta_{ijkl}^{AFM}$ have electro- and magnetostriction contributions, and as it was shown earlier [29], they have the following form: $\eta_{ijkl}^{FM} = -\eta_{ijkl} + q_{ijmn}^{(e)} s_{mnsp} q_{spkl}^{(m)} - \left(A_{ijsp} g_{ksn}^{(m)} g_{lpn}^{(m)} + B_{ijsp} g_{ksn}^{(e)} g_{lpn}^{(e)}\right)$ and $\eta_{ijkl}^{AFM} = \eta_{ijkl} + q_{ijmn}^{(e)} s_{mnsp} q_{spkl}^{(l)} - \left(C_{ijsp} g_{ksn}^{(m)} g_{lpn}^{(m)} + D_{ijsp} g_{ksn}^{(e)} g_{lpn}^{(e)}\right)$. Here $\eta_{ijkl}$ is the "bare" ME coupling tensor, $s_{mnsq}$ are elastic compliances, $g_{ijk}^{(e)}$ and $g_{ijk}^{(m)}$ are tensors of piezoelectric and piezomagnetic effects respectively. Since the electrostriction, magnitostriction, piezoelectric and piezomagnetic tensors strongly depend on the composition x, the ME coupling coefficients $\eta_{ijkl}^{FM}$ and $\eta_{ijkl}^{AFM}$ can vary essentially for PFN and PFT.

Variation of the thermodynamic potential (1)-(2) via polarization, magnetization, antiferromagnetic and antiferrodistortive order parameters in complex with Khalatnikov-type relaxation inclusion leads to the dynamic equations of state:

$$\Gamma_P \frac{\partial P_i}{\partial t} + \left(\alpha_P \delta_{il} - \eta_{mnil}^{FM} M_m M_n - \eta_{mnil}^{AFM} L_m L_n - \eta_{mnil}^{\Phi P} \Phi_m \Phi_n + q_{mnil}^{(e)} u_{mn}\right) P_l$$
$$+ \beta_{Pij} P_i P_j^2 = E_i^{ext} - \mu_{ji} M_j - \tilde{\mu}_{ji} L_j \tag{3a}$$

$$\Gamma_M \frac{\partial L_i}{\partial t} + \left(\alpha_L \delta_{il} - \eta_{ilmn}^{AFM} P_m P_n - \eta_{ilmn}^{\Phi L} \Phi_m \Phi_n + q_{mnil}^{(l)} u_{mn}\right) L_l + \beta_{Lij} L_i L_j^2 = -\tilde{\mu}_{ij} P_j, \tag{3b}$$

$$\Gamma_M \frac{\partial M_i}{\partial t} + \left(\alpha_M \delta_{il} - \eta_{ilmn}^{FM} P_m P_n - \eta_{ilmn}^{\Phi M} \Phi_m \Phi_n + q_{mnil}^{(m)} u_{mn}\right) M_l + \beta_{Mij} M_i M_j^2 = H_i^{ext} - \mu_{ij} P_j, \tag{3c}$$

$$\Gamma_\Phi \frac{\partial \Phi_i}{\partial t} + \left(\alpha_\Phi \delta_{il} - \eta_{ilmn}^{\Phi P} P_m P_n - \eta_{ilmn}^{\Phi M} M_m M_n - \eta_{ilmn}^{\Phi L} L_m L_n + q_{mnil}^{(r)} u_{mn}\right) \Phi_l + \beta_{\Phi ij} \Phi_i \Phi_j^2 = 0, \tag{3d}$$

Where $H_k^{ext}$ is an external magnetic field, $E_k^{ext}$ is an external electric field. Note, that the dynamic approach is needed to take into account the influence of domain structure motion on polarization or magnetization reversal.

Due to the presence of the strong bilinear ME coupling the magnetization induces a built-in electric field in Eq.(3a) and polarization induces a built-in magnetic field in Eq.(3c). In the case of linear response the built-in fields can be rewritten in the form:

$$E_i^{ME} = \mu_{ji} M_j \approx \mu_{ji}\left(M_j^S + \mu_0 \chi_{jk}^{(m)} H_k^{ext}\right), \tag{4a}$$

$$H_i^{ME} = \mu_{ij} P_j \approx \mu_{ij}\left(P_j^S + \varepsilon_0 \chi_{jk}^{(e)} E_k^{ext}\right). \tag{4b}$$

Hereinafter $M_j^S$ and $P_j^S$ are the spontaneous magnetization and polarization defined experimentally as remanent magnetization and polarization at zero external fields. Note that Eqs. (4) define the dependences of $E_i^{ME}$ and $H_i^{ME}$ on the composition $x$ and temperature $T$ via the dependence of polarization and magnetization on these quantities. In particular, if the component of the $E_i^{ME}$ conjugated with the spontaneous polarization component $P_i$ is higher than the critical field ($E_i^{cr}$) required for the domain wall motion, the polarization $P_i$ can be reversed by the field of appropriate direction. So that applied magnetic field can acts as the source of ferroelectric domain structure triggering observed by Evans et al.[8, 10].

The basic suggestion we will use below is the assumption of negligible influence of the magnetization and polarization on the pure antiferrodistortive order. In contrast the antiferrodistortive order parameter can significantly affect the polar and magnetic long range order [11]. The assumption is typical for classical multiferroics, such as BiFeO$_3$ [30, 31]. In order to describe the x-composition dependence of the antiferromagnetic (AFM) and ferromagnetic (FM) ordering we will use the approach [32] based on the percolation theory [33]. Also we assume a linear dependence of FM ordering on Fe content $x$ above the percolation threshold, $x = x_{cr}$. The critical concentration of percolation threshold $x_{cr}^F \approx 0.09$ [33] for the case of face-centered cubic sub-lattices of magnetic ions. The percolation threshold is supposed to be essentially higher for AFM ordering, $x_{cr}^A \approx 0.48$ (see e.g. [34, 35] and refs therein). Superscripts "*F*" and "*A*" in $x_{cr}^{F,A}$ designate the critical concentrations related to FM and AFM ordering respectively.

Using both of these assumptions we estimated the antiferromagnetic-paramagnetic (AFM-PM), ferromagnetic-paramagnetic (FM-PM) and ferroelectric-paraelectric (FE-PE) temperatures. For the composition $x > x_{cr}^A$, the temperature of the solid solution transition from the PM into AFM-ordering state is renormalized by the biquadratic ME and roto-antiferromagnetic couplings:

$$T_{PFY-PZT}^{AFM}(x) = T_Y^N \frac{(x - x_{cr}^A)}{(1 - x_{cr}^A)} - \frac{\eta_{AFM}(x)}{\alpha_{LT}} P_S^2(x) - \frac{\eta_{\Phi L}(x)}{\alpha_{LT}} \Phi_S^2(x). \tag{5}$$

Hereinafter Y=N for PFNx-PZT(1-x) or Y=T for PFTx-PZT(1-x). Measured Néel temperatures $T_Y^N$ for conventional bulk PFT and PFN are (143-170) K and (133-180) K correspondingly. $P_S$ and $\Phi_S$ are the spontaneous polarization and antiferrodistortive order parameter (e.g. oxygen octahedra tilt) correspondingly.

For the composition $x > x_{cr}^F$, the temperature of the solid solution transition from the PM into FM-ordering state is renormalized by the biquadratic ME and roto-antiferromagnetic couplings:

$$T_{PFY-PZT}^{FM}(x) = \theta_Y^C \frac{(x - x_{cr}^F)}{(1 - x_{cr}^F)} - \frac{\eta_{FM}(x)}{\alpha_{MT}} P_S^2(x) - \frac{\eta_{\Phi M}(x)}{\alpha_{MT}} \Phi_S^2(x). \tag{6}$$

The ferroelectric-paraelectric phase transition temperatures can be estimated from Eqs.(3a) as:

$$T_{PFY-PZT}^{FE} = x T_Y^C + (1-x) T_{PZT}^C - \frac{\eta_{\Phi P}(x)}{\alpha_{LT}} \Phi_S^2(T_{PFY-PZT}^{FE}, x). \tag{7}$$

The ferroelectric Curie temperatures $T_{PZT}^C = 666$ K [21], $T_{PFN}^C = (373 - 393)$ K and $T_{PFT}^C = (247 - 256)$ K were reliably and multiple times reproducibly defined for homogeneous bulk material with error margins depending on the sample preparation. In Eq.(5) we use the scalar approximation and regarded that the elastic stresses are absent.

As one can see from Eq.(7), without consideration of the possible anomalous tilt impact on the Curie temperature it was modeled using a *monotonic* linear law for the considered PZT-based solid solutions, $T_{PFY-PZT}^C(x) = x T_Y^C + (1-x) T_{PZT}^C$ [21]. The linear law is in a satisfactory agreement with experimental data shown in the figure 5 in 2013 of Sanchez et al [9], but the great scattering of the data was evident. However, Schiemer et al [11] experimentally studied the PFTx-PZT(1-x) multiferroic properties at x=0.4 and concluded that "the Raman data show a disappearance of all lines, compatible with a cubic transition, near 1123 K, but the polarization anomaly data suggest a higher temperature of approximately 1300 K". To clarify the situation, we will extract the compositional dependences of Curie temperature and Curie-Weiss constant from the temperature dependence of dielectric permittivity in the next section.

## III. Compositional dependences of Curie temperature, Curie-Weiss constants anharmonic stiffness, polarization and magnetization

The temperature dependences of relative dielectric permittivity of PFTx-PZT(1-x) and PFNx-PZT(1-x) ceramics are shown in the **Figures 1.** Symbols are experimental data from Sanchez et al [6, 9]. Solid curves are our fitting calculated using Curie-Weiss law with thermodynamic LGD model parameters listed in the **Tables 1-2**.

**Table 1.** LG coefficients for different compositions of PFNx-PZT(1-x)

| composition | x=0 | x=0.1 | x=0.2 | x=0.3 | x=0.4 | x=1 |
|---|---|---|---|---|---|---|
| $T_{CE}$ (K) | 666 | 668 | 670 | 590 | 620 | 375 |
| $C_{CE}$ (K) | $4.247 \times 10^5$ | $8 \times 10^5$ | $7.5 \times 10^5$ | $6.5 \times 10^5$ | $6.0 \times 10^5$ | $3.5 \times 10^5$ |
| $\beta$ * | $1.79 \times 10^8$ | $20 \times 10^8$ | $4 \times 10^8$ | $3 \times 10^8$ | $1.5 \times 10^8$ | $4 \times 10^8$ |
| $\gamma$ * | $8 \times 10^8$ | $10 \times 10^8$ | $10 \times 10^8$ | $90 \times 10^8$ | $20 \times 10^8$ | $20 \times 10^8$ |

**Table 2.** LG coefficients for different compositions of PFTx-PZT(1-x)

| composition | x=0 | x=0.3 | x=0.4 | x=1 |
|---|---|---|---|---|
| $T_{CE}$ (K) | 666 | 523 | 469 | 252 |
| $C_{CE}$ (K) | $4.247 \times 10^5$ | $2.9 \times 10^5$ | $3.3 \times 10^5$ | $2.5 \times 10^5$ |
| $\beta$ * | $1.79 \times 10^8$ | $3.5 \times 10^9$ | $2.3 \times 10^9$ | unknown* |
| $\gamma$ * | $8.0 \times 10^8$ | 0 | 0 | 0 |

It is seen from the **Figures 1a-c** that the maxima on the dielectric permittivity shifts to the lower temperatures with x increase from 0.3 to 0.4 for PFTx-PZT(1-x). As one can seen from the **Figure 1d** the maxima of the dielectric permittivity shifts to the lower temperatures with x increase from 0.2 to 0.3 and then slightly shifts to higher temperature for x=0.4 for PFNx-PZT(1-x). For all the solid solutions the dielectric anomaly transition temperature is within the range 400 – 700 K. Sanchez [9] reasonably attributed the anomaly as Curie temperature of FE phase transition.

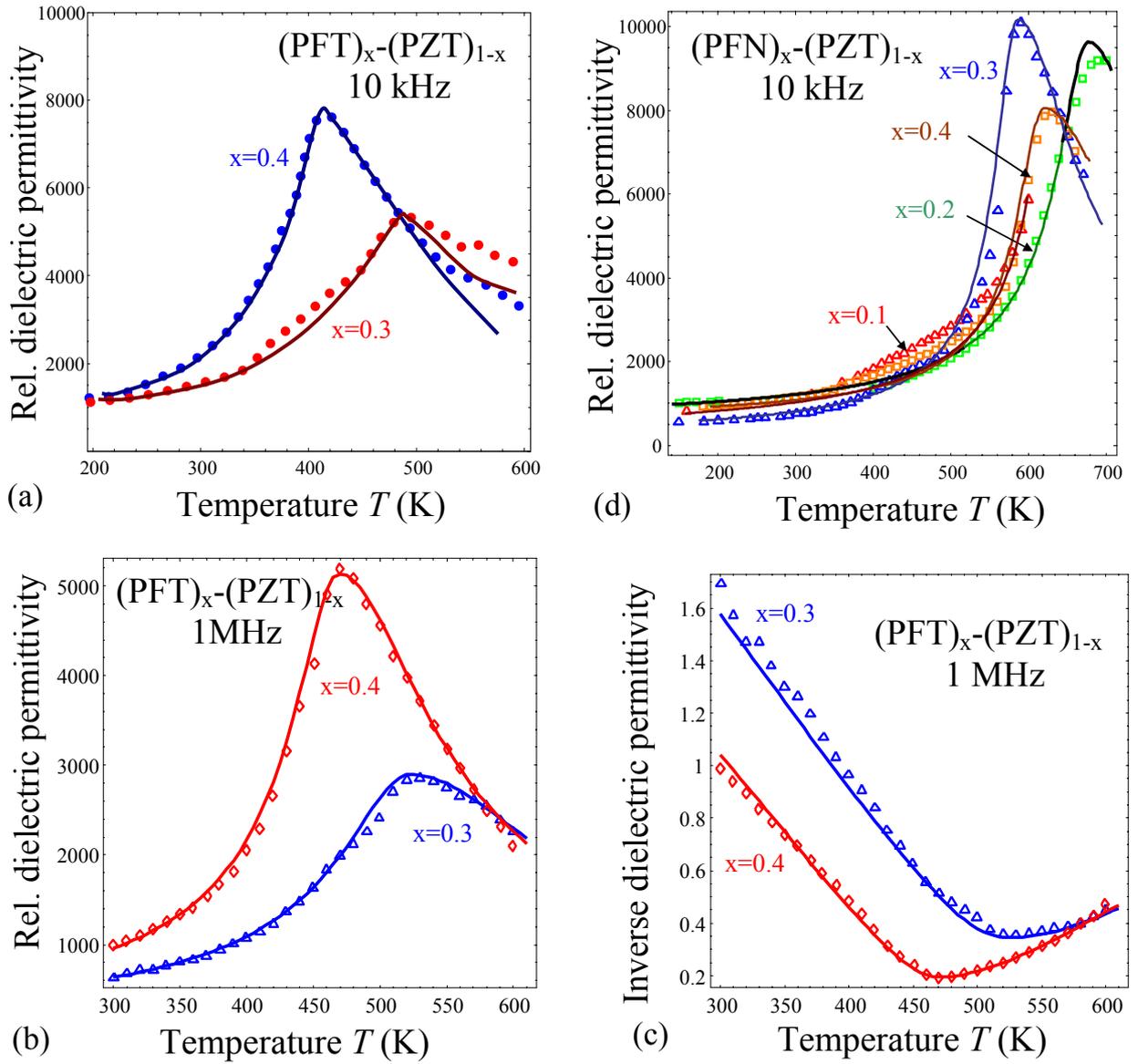

**Figure 1.** Temperature dependences of relative dielectric permittivity of PFTx-PZT(1-x) **(a-c)** and PFNx-PZT(1-x) **(d)** at 10 kHz and 1 MHZ. Symbols are experimental data from Sanchez et al [6, 9]. Solid curves are calculated using Curie-Weiss law within thermodynamic LGD model.

Composition dependences of the Curie temperatures and Curie-Weiss constants of PFTx-PZT(1-x) and PFNx-PZT(1-x) ceramics are shown in the **Figure 2**. Empty squires and triangles are data for PFTx-PZT(1-x) and PFNx-PZT(1-x) ceramics correspondingly extracted from the fitting of the dielectric permittivity temperature dependences shown in the **Figure 1** for x = 0 – 0.4. Circes on the plot (a) are the data from Sanchez et al [9] for x = 0.6 and 0.8. Dashed lines obey a linear composition laws $T^C_{PFY-PZT}(x) = xT^C_Y + (1-x)T^C_{PZT}$ and $C^{PFY-PZT}_{CW}(x) = xC^Y_{CW} + (1-x)C^{PZT}_{CW}$. Solid curves are spline-interpolation functions plotted using least squire method.

As one can see the linear laws satisfactory describe the compositional dependences of Curie temperatures and Curie-Weiss constants at compositions x > 0.4 (and sometimes for x=0.3 and 0.4 also), while the interpolation functions which describe adequately the dependences entire the composition range and evidently demonstrate the anomalous regions of the quantities nonlinear compositional dependences located at 0.1<x<0.4. At that the anomaly for PFNx-PZT(1-x) is much stronger than the one for PFTx-PZT(1-x). Actually, the interpolation functions for Curie temperature and Curie-Weiss constant of PFTx-PZT(1-x) quasi-monotonically sub-linearly decrease with the composition x increase at 0<x<1, indicating that the deviations from the linear-mixing laws $T^C_{PFT-PZT}(x) = xT^C_{PFT} + (1-x)T^C_{PZT}$ and $C^{PFT-PZT}_{CW}(x) = xC^{PFT}_{CW} + (1-x)C^{PZT}_{CW}$ are not very strong (compare solid curves and dashed lines in the **Figures 2**). At the same time the interpolation functions for Curie temperature have a smooth maxima and a very pronounced maxima for Curie-Weiss constant of PFNx-PZT(1-x) at compositions 0.1<x<0.4. The functions start to decrease quasi-linearly with the composition x increase only at 0.6<x<1, indicating that the deviations from the linear-mixing laws are very strong in the "anomalous" region of compositions 0.1<x<0.4 (compare solid curves and dashed lines in the **Figures 2**). To our opinion the anomalous may steam from the complex interplay of the polar and antiferrodistortive order parameters in the composition region.

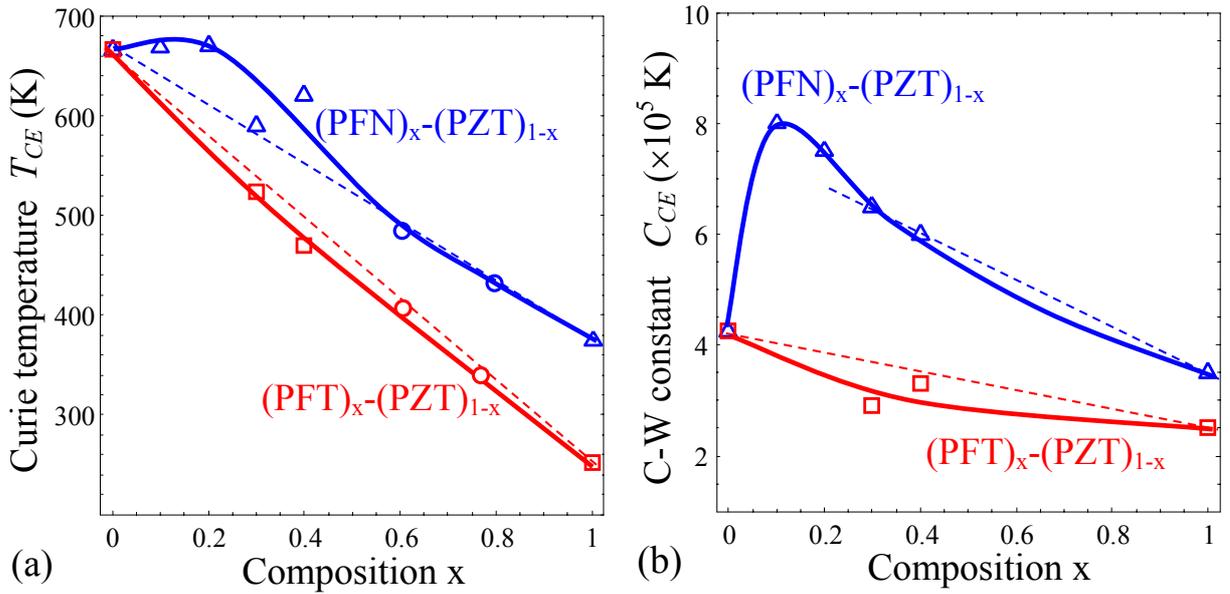

**Figure 2.** Composition dependences of the Curie temperatures (a) and Curie-Weiss constants (b) of PFTx-PZT(1-x) and PFNx-PZT(1-x) ceramics. Empty squires and triangles are data for PFTx-PZT(1-x) and PFNx-PZT(1-x) ceramics extracted at x = 0 – 0.4 from the fitting of the dielectric permittivity temperature dependences shown in the **Figure 1**. Corresponding LGD model parameters are listed in the **Tables 1-2**. Circes in the plot (a) are the data from Sanchez et al [9] for x = 0.6 and 0.8. Dashed lines obey a linear mixing law. Solid curves are spline-interpolation functions plotted using least squire method.

The most amazing fact is the strong composition dependences of the expansion coefficients β and γ, listed in the last lines of the **Tables 1** and **2** and shown in the **Figures 3a-b**. Note, that the anharmonic constants β and γ listed in the tables were determined from the temperature dependence dielectric permittivity and spontaneous polarization at room temperature measured by Sanchez et al [6, 9]. However, there is hardly possible to define reliably the anharmonic constants γ for PFTx-PZT(1-x), because the temperature dependences of the spontaneous polarization are unknown for the composition. At the same time the dielectric permittivity temperature dependence obey linear Curie-Weiss law $\varepsilon \sim \frac{C_{CW}}{T - T_C}$ with high accuracy. For the case we set the constant γ equal to zero.

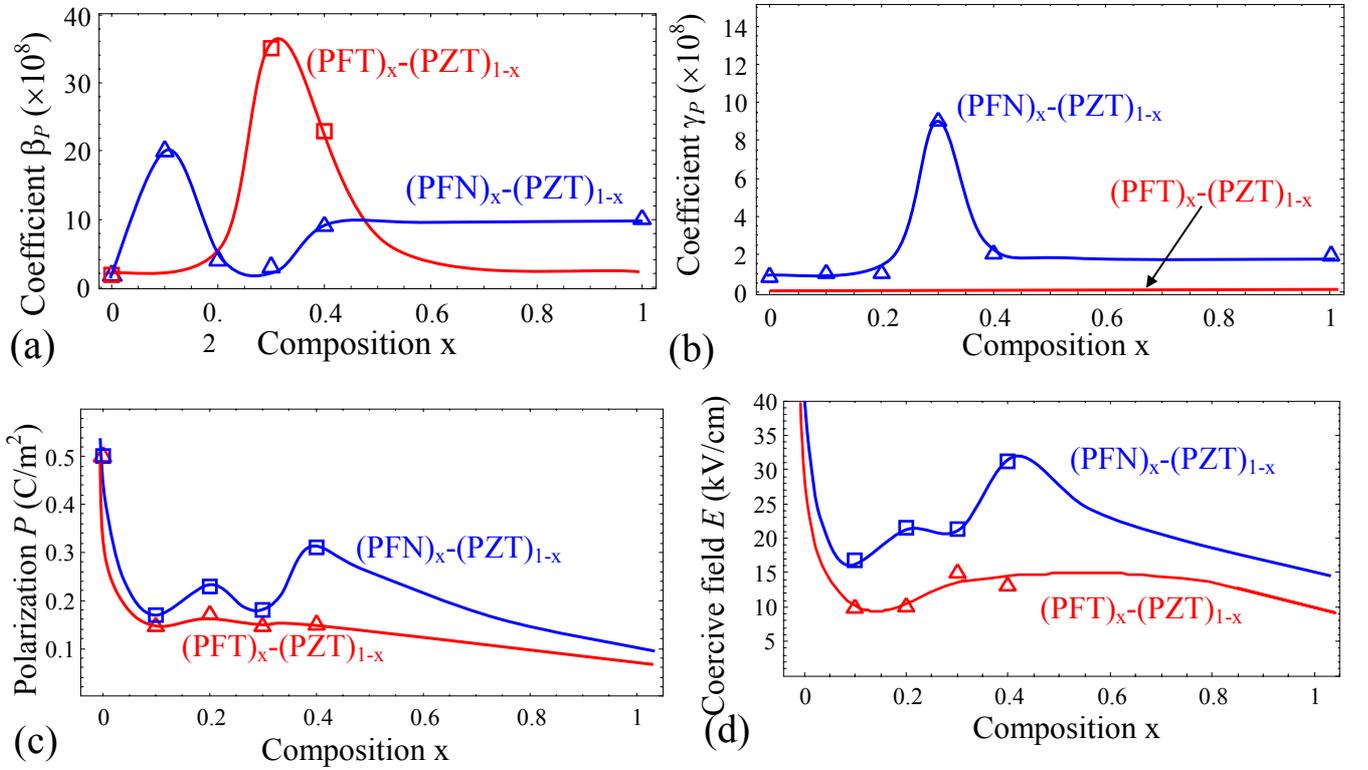

**Figure 3.** Dependences of the anharmonic constants of the ferroelectric part of the free energy $\beta_p$ **(a)** and $\gamma_p$ **(b),** remanent polarization **(c),** and coercive electric field **(d)** on the composition x of solid solutions PFTx-PZT(1-x) and PFNx-PZT(1-x) at room temperature. Boxes and triangles are data from Sanchez et al [6, 9]. Solid curves are spline-interpolation functions plotted using least squire method. Note that the anharmonic constants and remanent polarization values are correlated within our model.

To establish possible correlations between the structural, polar and magnetic properties, below we will discuss the composition dependences of the remanent polarization, magnetization and coercive fields with a special attention to the composition region 0.1<x<0.4.

**Figures 3c** and **3d** illustrates the anomalous (i.e. non-monotonic) composition x dependences of the remanent polarization and coercive electric field located in the composition range 0.1<x<0.4. The symbols indicate the values measured experimentally from hysteresis loops. In the static limit the remanent polarization should coincide with the spontaneous polarization. Interpolation functions fit the composition dependences of the remanent polarization within thermodynamic LGD model in a self-consistent way taking into account the parameters $\alpha_P$, $\beta_P$ and $\gamma_P$ defined from the temperature dependences of the dielectric susceptibility. The coercive fields were calculated within kinetic model indicating that the domain structure motion and retuning occur under the polarization reversal. There are anomalous oscillations of the points for polarization and coercive field. The anomaly is formally described by the strong composition dependences of the expansion coefficients $\beta_P$ and $\gamma_P$, which maxima and/or minima correlate with the anomalies of the remanent polarization.

Note, that high-temperature ferromagnetic properties have been observed exactly in the composition region 0.1<x<0.4 of PFTx-PZT(1-x) and PFNx-PZT(1-x) ceramics (see **Figure 4)**, where the dielectric measurements revealed anomalous composition behaviour of the Curie temperatures, Curie-Weiss constants, remanent polarization and coercive field (see **Figures 2-3)**. As one can see that the interpolation functions (solid curves) in the **Figures 4** fit well enough experimental points for remanent magnetization and magnetic coercive field. The non-monotonic dependence of the remanent magnetization is in agreement with the FM region boundary on the experimental phase diagram [6, 9]**.**

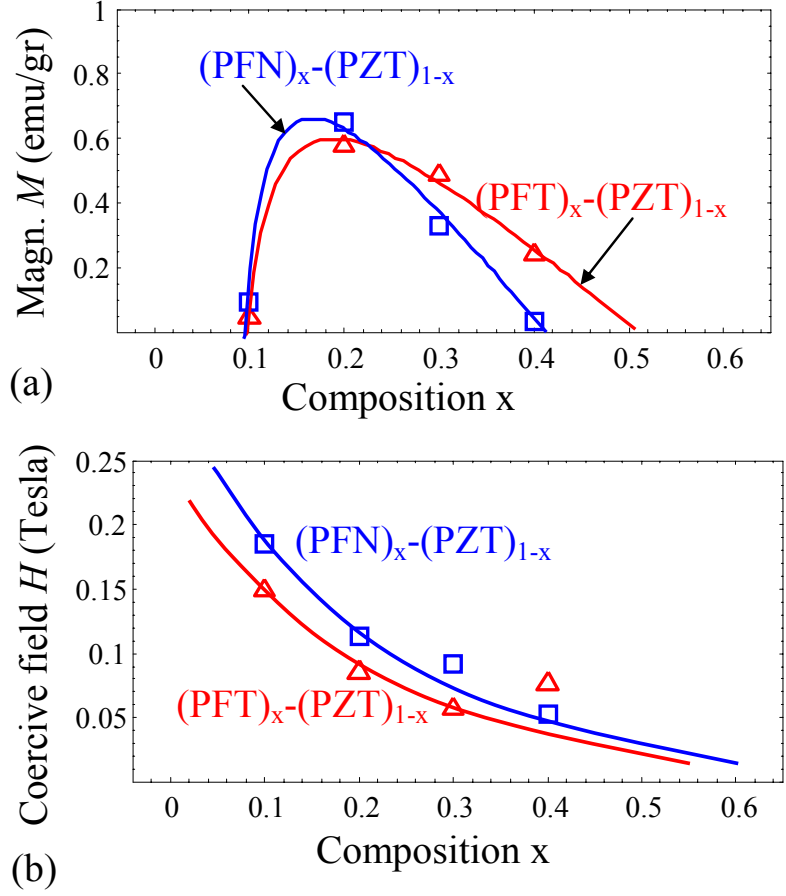

**Figure 4.** Remanent magnetization **(a)** and coercive magnetic field **(d)** dependences on the composition x of the solid solutions PFTx-PZT(1-x) and PFNx-PZT(1-x) at room temperature. Boxes and triangles are data from Sanchez et al [6, 9]. Solid curves are spline-interpolation functions plotted using least squire method.

### IV. Ferroelectric and ferromagnetic hysteresis loops: theory vs. experiment

The phenomenological consideration of ferroelectric systems dynamics is often based on macroscopic media theories, e.g. Landau–Ginzburg–Devonshire (LGD) theory [36]. It is general knowledge that the unique advantage is the possibility to obtain analytical expressions for considered physical quantities of the materials. The LGD theory applicability was shown to be correct for nanosized materials in the region, where macroscopic mean field models can be used [, 37]. The obtained analytical results allow to reach a comprehensive understanding of the physical process, as well as to predict and to analyze multi-scale size effects, which in turn opens an effective way to control and optimize nonlinear and hysteretic properties of materials. The mean field models are able to provide precise results only for the case of single phase materials. Rigorously speaking, the models are not appropriate for precise description of heterogeneous ferroics in the form of micro-grained ceramics with large amount of inter-grain defects.

However, it is possible to generate an "effective" model after averaging, which would provide suitable results [38, 39]. Physical grounds for the averaging in the case of PFTx-PZT(1-x) and PFNx-PZT(1-x) micro-ceramics originate from several sources: (a) defects accumulated by grain boundaries, which
12

create inhomogeneous stresses and electric fields; (b) different grain sizes. The defects and corresponding internal disorder can lead to the renormalization of LGD model parameters, primarily $\alpha_P$, $\beta_P$ and $\alpha_M$, $\beta_M$. Below the averaging of polarization and magnetization hysteresis dynamics calculated from Eqs.(3) is performed over the aforementioned parameters using the same normal Gaussian distribution for all compositions x. The averaging leads to the noticeable smearing and tilt of the loop shape in agreement with Sanchez et al experimental data [6, 9].

The calculated ferroelectric hysteresis loops are shown in the **Figures 5a** and **5b** for several compositions x. The loops were calculated within thermodynamic LGD model with the averaging of the loops over the expansion coefficient $\alpha_P$ and $\beta_P$. The averaged values of the remanent polarization and coercive electric field were fixed in agreement with the **Figures 3c** and **3d**. This was the only reasonable way to fit the non-monotonic dependence of the remanent polarization on the composition x. The reference coercive field $E_c^{max}$ was chosen as for the homogeneous infinite material (dotted loop). We remarkably note the semi-quantitative agreement between the shape of calculated loops and remanent polarization value with the loops measured by Sanchez et al (compare the present **Figure 4a** with the figure 5 in the ref [6], and the present **Figure 4b** with the figure 3b in the ref [9]).

The calculated ferromagnetic hysteresis loops is shown in the **Figures 5c** and **5d** for several compositions x. The loops were calculated within LGD model with the averaging of the loops over the expansion coefficient $\alpha_M$ and $\beta_M$. The averaged values of the remanent magnetization and coercive magnetic field were fixed in agreement with the **Figures 4a** and **4b**. In particular, the nonmonotonic dependence of the remanent magnetization is in agreement with the **Figure 4a.** The reference coercive field $H_c^{max}$ was chosen as for the hypothetic ferromagnetic material. We would like to underline on the reasonable agreement between the shape of calculated loops and remanent magnetization value with the loops measured by Sanchez et al (compare the present **Figure 4c** with the figure 7 in the ref [6], and the present **Figure 4d** with the figure 4a in the ref [9]).

However it appeared hardly possible to fit quantitatively the experimental values of coercive electric ($E_c \sim (10 - 35)$ kV/cm) and magnetic ($H_c \sim (0.05 - 0.2)$ Tesla) fields [6, 9]. Within the LGD model it happens with "rigidly" fixed several expansion coefficients $\alpha$, $\beta$, which give the values of coercive fields ($E_c^{FE} = 2\sqrt{-\alpha_P^3/27\beta_P}$ and $E_c^{FM} = 2\sqrt{-\alpha_M^3/27\beta_M}$ for the second order phase transition [40]), spontaneous polarization and magnetization ($P_S = \sqrt{-\alpha_P/\beta_P}$ and $M_S = \sqrt{-\alpha_M/\beta_M}$), dielectric and magnetic susceptibilities ($\chi_E \sim \alpha_P^{-1}$ and $\chi_M \sim \alpha_M^{-1}$ in PE phase). However, if one fits the polarization and dielectric susceptibility well, the coercive field cannot be fitted independently. Therefore the modelling in the framework of thermodynamic LGD approach usually results in very large values of the coercive field, which much larger (up to several orders of magnitude) than those observed in experiments for inhomogeneous ferroics, in particular for ceramics with inherent numerous defects located at the grain boundaries. The impossibility to fit the experimental loops of polarization and magnetization reversal in the studies of ferroelectric micro-ceramics proves that the polarization reversal process is the limited here by domain nucleation and growth.

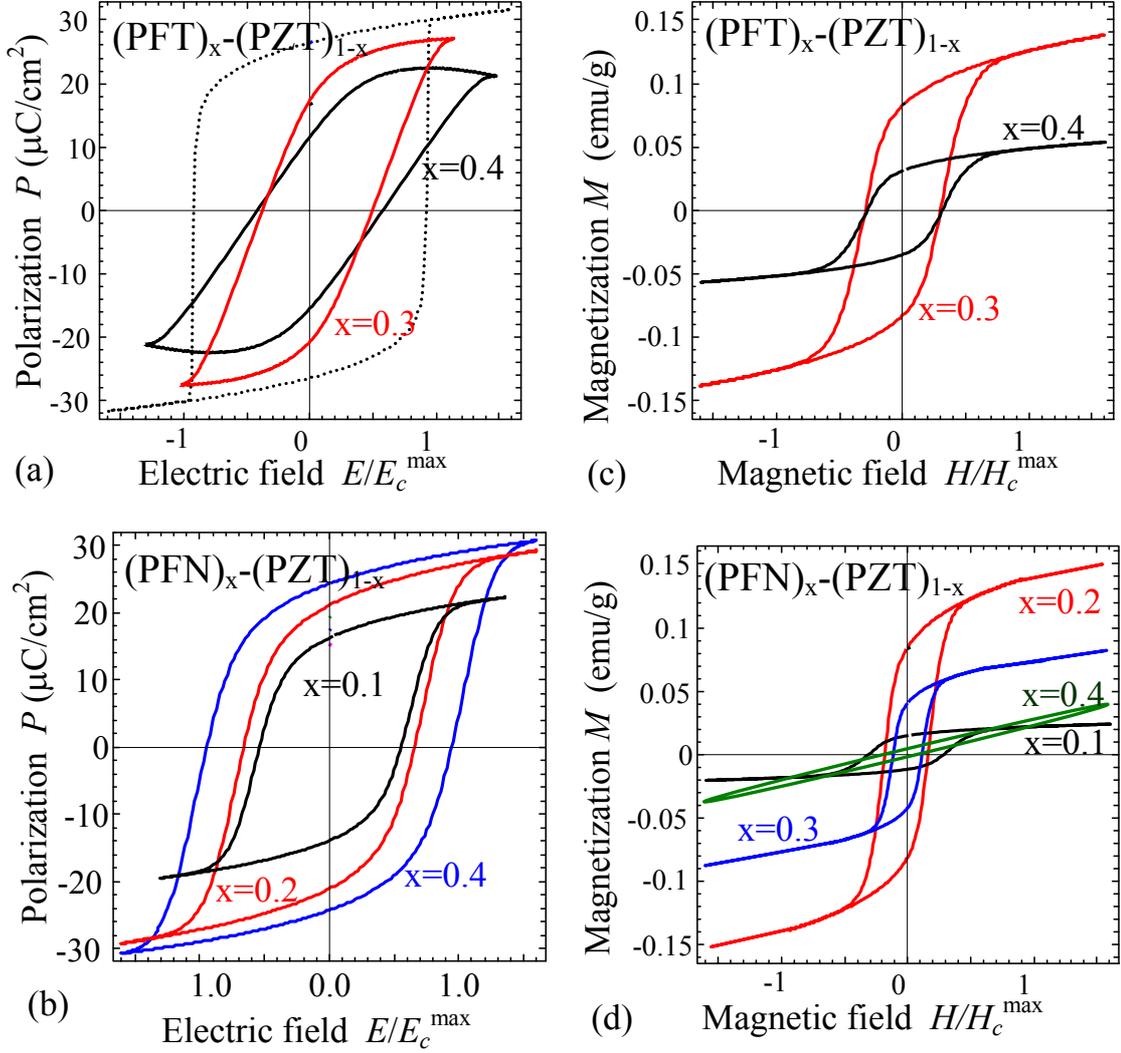

**Figure 5.** Calculated ferroelectric **(a,b)** and ferromagnetic **(c,d)** hysteresis loops of PFTx-PZT(1-x) **(a,c)** and PFNx-PZT(1-x) **(b,d)** at room temperature.

**Figure 6a** shows anomalous changes of ferroelectric hysteresis loop shape under the temperature increase from 300 K to 550 K in PFTx-PZT(1-x) (x=0.3). **Figure 6b** shows the anomalous temperature enhancement of the remanent polarization extracted from the loops. Symbols are experimental data extracted from the loops presented in the figure 6 of Sanchez et al 2011 [6], solid curve is our interpolation function $P_S(T) \approx P_{S0} + A_T(T - T_O)^3$ with parameters $P_{S0} = 17$ μC/cm$^2$, $A_T = 1.1 \times 10^6$ μC/(K$^3 \times$cm$^2$) and $T_O = 250$ K. The question is whether the function can be in agreement with our theoretical model allowing for the impact on the structural order parameter with specific temperature dependence.

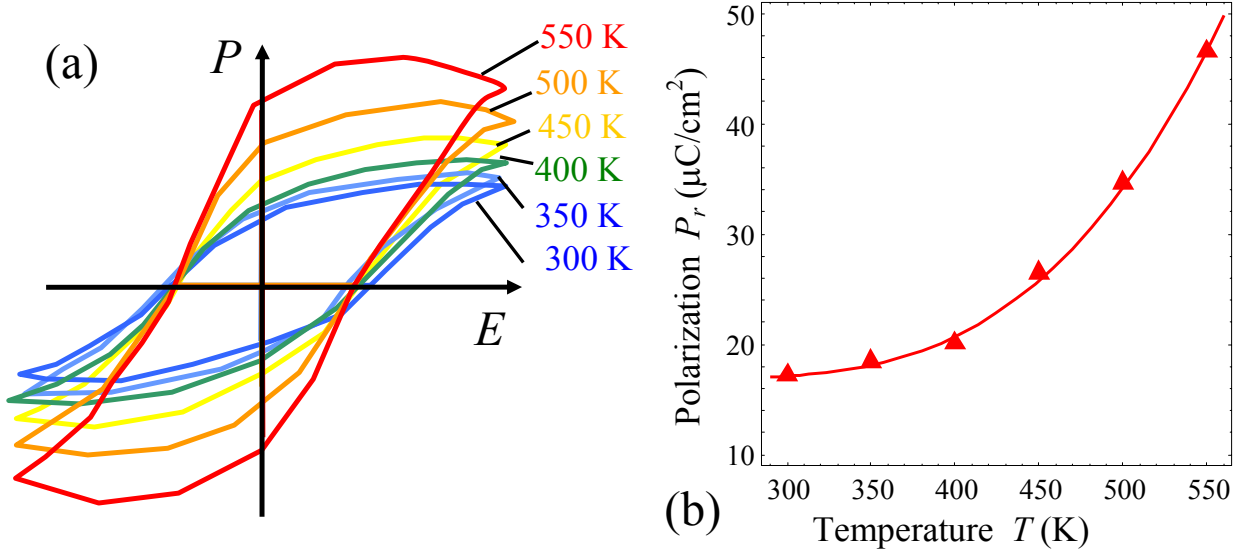

**Figure 6. (a)** Ferroelectric hysteresis loop shape for PFTx-PZT(1-x) (x=0.3) measured at different temperatures 300, 350, 400, 450, 500, 550 K by Sanchez et al 2011 [6]. **(b) Anomalous temperature dependence of the remanent polarization.** Symbols are experimental data extracted from the loops **(a)** as observed by Sanchez et al 2011 [6], solid curve is our interpolation function $P_S(T) \approx P_{S0} + A_T(T - T_O)^3$ with parameters $P_{S0}$ = 17 μC/cm$^2$, $A_T = 1.1 \times 10^6$ μC/(K$^3$·cm$^2$) and $T_O$ = 250 K.

Note, that Sanchez et al [6o9] provided the following explanation of the anomalous polarization behaviour. They underlined that above the ferroelectric phase transition of PFTx-PZT(1-x) (x=0.3), namely above 520 K, the enhancement in polarization value confirms the structural orthorhombic-tetragonal **(O-T)** phase transition and rules out a higher symmetry cubic phase with zero polarization. High- and low-temperature Raman studies revealed two more structural phase transitions, one is the tetragonal-cubic **(T-C)** at higher temperature (> 1300 K) and the other is orthorhombic-rhombohedral **(O-R)** near 250 K. The temperature evolution of the polarization favours an orthorhombic to tetragonal phase transition in PFTx-PZT(1-x) and PFNx-PZT(1-x). Actually, we obtained the fitting value of $T_O$ = 250 K that indicates the **O-R** transition. It is not excluded that the coexistence of several phase transitions in the considered temperature region may be the main source of the specific form of interpolation function for $P_S(T)$.

To resume, **Figures 5-6** show the room and higher temperature well saturated polarization-electric field and magnetization-magnetic field hysteresis loop of PFTx-PZT(1-x) and PFNx-PZT(1-x) ceramics in agreement with experiments [6, 9].

### V. Electric field control of magnetization due to magnetoelectric effect

The experimental dependences of the remanent magnetization shown in the left-hand side inset of the figure 9 from Sanchez et al [6] allows to estimate the relative contribution of bilinear (μ$^*$) and biquadratic ($\eta^*$) magnetoelectric effect. Namely, we extracted the remanent magnetization value $M_S \approx 0.01531$ emu at zero electric field, $M_S \approx 0.01517$ emu at positive electric field $E=35$ kV/cm and $M_S \approx 0.0151$ emu at negative electric field $-35$ kV/cm from the top, middle and bottom straight lines respectively in the abovementioned inset. Then using the simple expansion

$$M_S(E) \approx M_S(0) + \mu^* E + \eta^* E^2 \tag{8}$$

we obtained the values of bilinear and biquadratic ME coupling coefficients, $\mu^* = 1.00 \times 10^{-7}$ emu·m/V and $\eta^* = -1.43 \times 10^{-9}$ emu·m$^2$/V$^2$. The obtained values of both linear and biquadratic ME coupling coefficients are high enough for single phase materials.

The change in remanent magnetization under external electric field ($M_S(E)$) is shown in the **Figure 7a.** Symbols are experimental data [6] corresponding to the direct measurements of $M_S(E)$, dashed parabolic curve is calculated using Eq.(8) with parameters $\mu^* = +1.00 \times 10^{-7}$ emu×m/V and $\eta^* = -1.43 \times 10^{-9}$ emu×m$^2$/V$^2$; solid parabolic curve is calculated using Eq.(8) with the parameters $\mu^* = -1.00 \times 10^{-7}$ emu×m/V and $\eta^* = -1.43 \times 10^{-9}$ emu×m$^2$/V$^2$, corresponding to the best fitting on the base of the least squire method.

By comparison of the ME coupling parameters corresponding to the dashed and solid parabolic curves we can conclude that they correspond to the same $\eta^*$ and $|\mu^*|$ values, but to different signs of the linear ME coefficient $\mu^*$. Thus that there is a disagreement in the right-hand side inset in the figure 9 ref.[6], where $M_S(-E)$ is more than $M_S(+E)$ contrary to the left-hand side inset fig.6, where $\mu^* > 0$.

**Figure 7b** demonstrates the linear ME effect contribution into the remanent magnetization, $M_S(+E) - M_S(-E) \approx \mu^* E$. The positive sign of $\mu^*$ for dashed straight line and the negative sign of $\mu^*$ for solid straight line are pretty obvious. To our mind there may be a discrepancy, namely muddled up labels for $E=35$ kV/cm and $E=-35$ kV/cm, in the experimental data represented in the right-hand side inset of fig.9 [6]. Because of this only biquadratic ME coupling coefficient $\eta^*$ can be defined correctly from the experiment [6]. Note that the coefficient is negative, and using Eq.(21) from Ref.[29], one can suggest the same sign of the ME susceptibility, $\chi_{ME} = \dfrac{\eta^* \sqrt{\chi_E \chi_M}}{2\sqrt{\beta_P \beta_M}}$, where the dielectric and magnetic susceptibilities are $\chi_E = \dfrac{C_E}{(T_{CE}^* - T)}$ and $\chi_M = \dfrac{C_M}{(T_{CM}^* - T)}$ correspondingly.

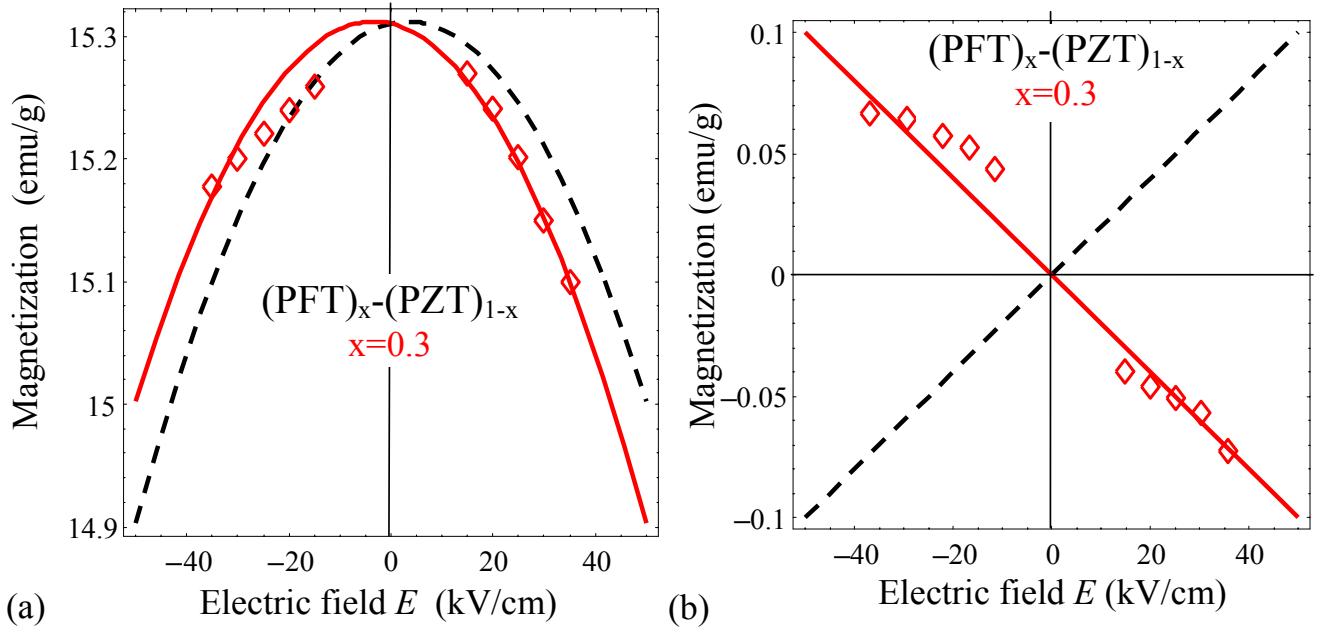

**Figure 7. (a)** The dependence of magnetization vs. electric field at room temperature. Symbols are experimental data [6], solid curve is calculated using Eq.(8) with parameters $\mu^* = +1.00 \times 10^{-7}$ emu·m/V and $\eta^* = -1.43 \times 10^{-9}$ emu·m$^2$/V$^2$; dashed curve is calculated using Eq.(8) with the best fitting parameters $\mu^* = -1.00 \times 10^{-7}$ emu·m/V and $\eta^* = -1.43 \times 10^{-9}$ emu·m$^2$/V$^2$. **(b)** Demonstration of the linear ME effect contribution into the remanent magnetization, $M_S(+E) - M_S(-E) \approx \mu^* E$.

Note that the positive sign of ME coupling coefficient one can find in the page 12 of Ref.[6] and it is agreement with the results of Evans [8, 10]. Similar ME effect was observed for x=0.4 by Sanchez et al 2011 [6] and can be readily calculated with experimental results in hands.

## VII. Conclusion

Our consideration was performed on the basis of phenomenological LGD theory to obtain analytical expressions for physical properties of multiferroic PFT$_x$PZT$_{1-x}$ and PFN$_x$PZT$_{1-x}$ microceramics. In particular we performed calculations of temperature, composition and external fields dependences of ferroelectric, ferromagnetic and antiferromagnetic phases transition temperatures, remanent polarization, magnetization, hysteresis loops, coercive fields, dielectric permittivity and magnetoelectric coupling. Special attention was paid to comparison of the developed theory with experiment. It appeared possible to describe pretty good main experimental results observed by Sanchez et al. [6,9]. There is reasonable agreement between the shape of calculated hysteresis loops and remnant polarization value with loops and polarization measured by Sanchez et al. [6,9] (see Figs. 4 and 5a). Information about linear and nonlinear ME coupling coefficients was extracted from the experimental data [6] for (PFT)$_{0.3}$(PZT)$_{0.7}$. From the fitting of experimental data with theoretical formula it appeared possible to obtain composition dependence of Curie-Weiss constant (Fig.2a-2b) that is known to be inversely proportional to harmonic (linear) dielectric stiffness $\alpha_P$ of the free energy ferroelectric part, as well as the strong nonlinear dependence of anharmonic parameter $\beta_P$ (see Fig.3a-3b). Keeping in mind the essential influence of these parameters on the multiferroic properties the obtained results open the way to govern practically all the material properties with the help of suitable composition choice. The forecast of the strong enough influence of antiferrodistortive order parameter on the transition temperatures and so on the phase diagrams and properties of multiferroics is made on the basis of the developed theory.

Allowing for Schiemer et al [11] recently observed several structural phase transitions in $(PFT)_{0.4}(PZT)_{0.6}$ and underlined their similarity to those in $BaTiO_3$ as well as in $K_{1-x}Na_xNbO_3$ with antiferrodistorsive phase transition [41] we included antiferrodistortive order parameter in consideration. Unfortunately to the best of our knowledge nothing is known now about it in PFT–PZT, so that we consider our formula (5)–(7) as a theoretical forecast.

## Acknowledgments

Authors acknowledge National Academy of Sciences of Ukraine and STCU.

**References**

[1] N. A. Spaldin, and R. Ramesh, Electric-field control of magnetism in complex oxide thin films. MRS Bull. 33, 1047 (2008).

[2] J. F. Scott, Applications of magnetoelectrics. J. Mater. Chem. 22, 4567–4574 (2012).

[3] A. Pyatakov, P. Zvezdin, A. K. Magnetoelectric and multiferroic media. Physics—Uspekhi 55, 557–581 (2012).

[4] R. O. Cherifi, V. Ivanovskaya, L. C. Phillips, A. Zobelli, I. C. Infante, E. Jacquet, V. Garcia, S. Fusil, P. R. Briddon, N. Guiblin, A. Mougin, A. A. Ünal, F. Kronast, S. Valencia, B. Dkhil, A. Barthélémy, M. Bibes. Electric-field control of magnetic order above room temperature, *Nature Materials,* 13, 345–351 (2014).

[5] James F Scott. "Room-temperature multiferroic magnetoelectrics." NPG Asia Materials 5, no. 11: e72 (2013).

[6] Dilsom A. Sanchez, N. Ortega, Ashok Kumar, R. Roque-Malherbe, R. Polanco, J. F. Scott, and Ram S. Katiyar. Symmetries and multiferroic properties of novel room-temperature magnetoelectrics: Lead iron tantalate–lead zirconate titanate (PFT/PZT)." AIP Advances 1, 042169 (2011).

[7] L.W. Martin, R. Ramesh, Multiferroic and magnetoelectric heterostructures. Acta Materialia, **60**, 2449-2470 (2012)

[8] D.M. Evans, A. Schilling, Ashok Kumar, D. Sanchez, N. Ortega, M. Arredondo, R.S. Katiyar, J.M. Gregg & J.F. Scott. «Magnetic switching of ferroelectric domains at room temperature in multiferroic PZTFT»Nature Communications, 4, 1534 (2013).

[9] Dilsom A. Sanchez, Nora Ortega, Ashok Kumar, G. Sreenivasulu, Ram S. Katiyar, J. F. Scott, Donald M. Evans, Miryam Arredondo-Arechavala, A. Schilling, and J. M. Gregg. "Room-temperature single phase multiferroic magnetoelectrics: Pb (Fe, M) x (Zr, Ti)(1− x) O3 [M= Ta, Nb]."J. Appl. Phys. 113, 074105 (2013).

[10] D. M. Evans, A. Schilling, Ashok Kumar, D. Sanchez, N. Ortega, R. S. Katiyar, J. F. Scott and J. M. Gregg, "Switching ferroelectric domain configurations using both electric and magnetic fields in Pb (Zr, Ti) O3–Pb (Fe, Ta) O3 single-crystal lamellae." Phil. Trans. R. Soc. A 372, 20120450 (2014)

[11] J. Schiemer, M. A. Carpenter , D. M. Evans, J. M. Gregg, A. Schilling, M. Arredondo, M. Alexe, D. Sanchez, N. Ortega , R. S. Katiyar, M. Echizen, E. Colliver, S. Dutton, and J. F. Scott. "Studies of the Room-Temperature Multiferroic Pb (Fe0. 5Ta0. 5) 0.4 (Zr0. 53Ti0. 47) 0.6 O3: Resonant Ultrasound Spectroscopy, Dielectric, and Magnetic Phenomena." Adv. Funct. Mater., **24**, 2993–3002 (2014)

[12] Sergey A Ivanov, Roland Tellgren, Häkan Rundlof, Noel W. Thomas, and Supon Ananta. "Investigation of the structure of the relaxor ferroelectric Pb (Fe1/2Nb1/2) O3 by neutron powder diffraction." Journal of Physics: Condensed Matter 12, no. 11: 2393 (2000).

[13] Andrea Falqui, Nathascia Lampis, Alessandra Geddo-Lehmann, and Gabriella Pinna. «Low-Temperature Magnetic Behavior of Perovskite Compounds $PbFe_{1/2}Ta_{1/2}O_3$ and $PbFe_{1/2}Nb_{1/2}O_3$» J. Phys. Chem. **109**, 22967-22970 (2005).

[14] R.K. Mishra, R.N.P. Choudhary, and A. Banerjee. "Bulk permittivity, low frequency relaxation and the magnetic properties of Pb (Fe1/2Nb1/2) O3 ceramics." Journal of Physics: Condensed Matter 22, 2: 025901 (2010).

[15] W. Kleemann, V. V. Shvartsman, P. Borisov, and A. Kania. "Coexistence of antiferromagnetic and spin cluster glass order in the magnetoelectric relaxor multiferroic PbFe0.5Nb0.5O3." Phys. Rev. Lett. 105, no. 25: 257202 (2010).

[16] M. H. Lente, J. D. S. Guerra, G. K. S. de Souza, B. M. Fraygola, C. F. V. Raigoza, D. Garcia, and J. A. Eiras. "Nature of the magnetoelectric coupling in multiferroic Pb (Fe 1/2 Nb 1/2) O 3 ceramics." Phys. Rev. B 78, 054109 (2008).

[17] V. V. Bhat, A. M. Umarji, V. B. Shenoy, and U. V. Waghmare. "Diffuse ferroelectric phase transitions in Pb-substituted Pb Fe 1∕2 Nb 1∕2 O 3." Physical Review B 72, no. 1: 014104 (2005).

[18] Chandrahas Bharti, Alo Dutta, Santiranjan Shannigrahi, T.P. Sinha. "Electronic structure, magnetic and electrical properties of multiferroic PbFe1/2Ta1/2O3." Journal of Magnetism and Magnetic Materials 324, 955–960 (2012).

[19] Nathascia Lampis, Cesare Franchini, Guido Satta, Alessandra Geddo-Lehmann, and Sandro Massidda. "Electronic structure of PbFe 1/2 Ta 1/2 O 3: Crystallographic ordering and magnetic properties." Phys. Rev. B 69, 064412 (2004).

[20] W.Z. Zhu, A. Kholkin, P.Q. Mantas, J.L. Baptista, " "Preparation and characterisation of Pb (Fe 1/2 Ta 1/2) O 3 relaxor ferroelectric." " J. Eur. Ceram. Soc. 20, 2029–2034 (2000).

[21] M. J. Haun, Z. Q. Zhuanga, E. Furman, S. J. Jang & L. E. Cross. " Thermodynamic theory of the lead zirconate-titanate solid solution system, Part III: Curie constant and sixth-order polarization interaction dielectric stiffness coefficients" Ferroelectrics, **99**, 45-54 (1989).

[22] Maya D. Glinchuk, Eugene A. Eliseev, Anna N. Morozovska, New room temperature multiferroics on the base of single-phase nanostructured perovskites. J. Appl. Phys. **116**, 054101 (2014)

[23] Nawnit Kumar, Avijit Ghosh, R.N.P. Choudhary. Electrical behavior of Pb(Zr0.52Ti0.48)0.5(Fe0.5Nb0.5)0.5O3 ceramics. Materials Chemistry and Physics, 130, 381–386 (2011).

[24] M.D. Glinchuk, A.N. Morozovska, E.A. Eliseev, and R. Blinc. Misfit strain induced magnetoelectric coupling in thin ferroic films. J. Appl. Phys. **105**, 084108 (2009).

[25] Eugene A. Eliseev, Maya D. Glinchuk, Venkatraman Gopalan, Anna N. Morozovska. Rotomagnetic couplings influence on the magnetic properties of antiferrodistortive antiferromagnets http://arxiv.org/abs/1409.7108

[26] M.A. Carpenter, A.I. Becerro and Friedrich Seifert, "Strain analysis of phase transitions in (Ca, Sr) TiO3 perovskites." American Mineralogist, 86, 348–363 (2001).

[27] Yijia Gu, Karin Rabe, Eric Bousquet, Venkatraman Gopalan, and Long-Qing Chen. "Phenomenological thermodynamic potential for CaTiO 3 single crystals." Phys. Rev. **B 85**, 064117 (2012).

[28] Eugene A. Eliseev, Anna N. Morozovska, Yijia Gu, Albina Y. Borisevich, Long-Qing Chen and Venkatraman Gopalan, and Sergei V. Kalinin. "Conductivity of twin-domain-wall/surface junctions in ferroelastics: Interplay of deformation potential, octahedral rotations, improper ferroelectricity, and flexoelectric coupling." Phys. Rev. **B 86**, 085416 (2012)

[29] M.D. Glinchuk, E.A. Eliseev, A.N. Morozovska, R. Blinc. Giant magnetoelectric effect induced by intrinsic surface stress in ferroic nanorods. Phys. Rev. **B 77**, 024106 (2008).

[30] G. Catalan and James F. Scott. Adv. Mater. **21**, 1–23 (2009).

[31] G. Catalan, J. Seidel, R. Ramesh, and J. F. Scott. Rev. Mod. Phys. **84**, 119 (2012).

[32] Eugene A. Eliseev, Maya D. Glinchuk, Victoria V. Khist, Chan-Woo Lee, Chaitanya S. Deo, Rakesh K. Behera, and Anna N. Morozovska. "New multiferroics based on $Eu_xSr_{1-x}TiO_3$ nanotubes and nanowires." J. Appl. Phys. **113**, 024107 (2013)

[33] B.I. Shklovskii and A.L. Efros. Electronic properties of doped semiconductors. (Springer-Verlag, Berlin 1984). 388 *Pages*.

[34] J. Adler, R. G. Palmer, and H. Meyer, " Transmission of order in some unusual dilute systems" Phys. Rev. Lett. 58, 882 (1987).

[35] Harry Fried and M. Schick. "Nonlocal percolation in an antiferromagnetic Potts model." Phys. Rev. **B 38**, 954–956 (1988)

[36] V. M. Fridkin, S. Ducharme, "Ferroelectricity at the nanoscale", UFN, 184:6 (2014), 645–651.

[37] A.N. Morozovska, M.D. Glinchuk, E.A. Eliseev. "Phase transitions induced by confinement of ferroic nanoparticles" Physical Review **B 76**, 014102-1-13 (2007).

[38] Alexander K. Tagantsev, Igor Stolichnov, and Nava Setter, Jeffrey S. Cross and Mineharu Tsukada, "Non-Kolmogorov-Avrami switching kinetics in ferroelectric thin films." Phys. Rev. B 66, 214109 (2002).

[39] A.N. Morozovska, E.A. Eliseev, "Strain-Induced Disorder in Ferroic Nanocomposites" in: K.D. Sattler (Ed.), Handbook of Nanophysics, Functional Nanomaterials, Taylor&Francis (2010), Vol.5.

[40] L.D. Landau, E.M. Lifshitz, L. P. Pitaevskii. Electrodynamics of Continuous Media, (Second Edition, Butterworth-Heinemann, Oxford, 1984).

[41] M. Ahtee, and A.M. Glazer, "Lattice parameters and tilted octahedra in sodium-potassium niobate solid solutions." *Acta Crystallogr. Sect. A* **32**(3), 434–446 (1976).